\documentclass[aps,prl,reprint,superscriptaddress,nofootinbib]{revtex4-2}
\usepackage[utf8]{inputenc}
\usepackage{amsfonts}
\usepackage{graphicx}% Include figure files
\usepackage{color}
\usepackage{soul} %needed for strikeout
\usepackage{amsmath}
\usepackage{hyperref}
\usepackage{braket}
\usepackage{mathtools}
\usepackage{lipsum}
\hypersetup{colorlinks=true, citecolor=cyan, urlcolor=blue, linkcolor=blue}
\usepackage{mathrsfs}

\newcommand{\RMP}[4]{\textit{#1}, Rev. Mod. Phys. \textbf{#2}, #3 (#4)}

\newcommand{\PRL}[4]{\textit{#1}, Phys. Rev. Lett. \textbf{#2}, #3 (#4)}
\newcommand{\PRR}[4]{\textit{#1}, Phys. Rev. Research. \textbf{#2}, #3 (#4)}
\newcommand{\PRA}[4]{\textit{#1}, Phys. Rev. A \textbf{#2}, #3 (#4)}
\newcommand{\PRB}[4]{\textit{#1}, Phys. Rev. B \textbf{#2}, #3 (#4)}

\newcommand{\PRApplied}[4]{\textit{#1}, Phys. Rev. Appl. \textbf{#2}, #3 (#4)}

\newcommand{\Science}[4]{\textit{#1}, Science \textbf{#2}, #3 (#4)}
\newcommand{\Nature}[4]{\textit{#1}, Nature \textbf{#2}, #3 (#4)}
\newcommand{\NPhys}[4]{\textit{#1}, Nat. Phys \textbf{#2}, #3 (#4)}
\newcommand{\NComm}[4]{\textit{#1}, Nat. Comm \textbf{#2}, #3 (#4)}
\newcommand{\CommPhys}[4]{\textit{#1}, Comm. Phys \textbf{#2}, #3 (#4)}

\newcommand{\NanoLett}[4]{\textit{#1}, Nano Lett. \textbf{#2}, #3 (#4)}

\newcommand{\APL}[4]{\textit{#1}, Appl. Phys. Lett. \textbf{#2}, #3 (#4)}

\newcommand{\PhysRep}[4]{\textit{#1}, Phys. Rep. \textbf{#2}, #3 (#4)}
\newcommand{\NNano}[4]{\textit{#1}, Nat. Nanotechnol. \textbf{#2}, #3 (#4)}
\newcommand{\NatComm}[4]{\textit{#1}, Nat. Commun. \textbf{#2}, #3 (#4)}

\newcommand{\arXiv}[3]{\textit{#1}, arXiv preprint arXiv:#2 (#3)}
\newcommand{\JAP}[4]{\textit{#1}, J. Appl. Phys. \textbf{#2}, #3 (#4)}
\newcommand{\JCP}[4]{\textit{#1}, J. Chem. Phys. \textbf{#2}, #3 (#4)}
\newcommand{\PRMaterials}[4]{\textit{#1}, Phys. Rev. Materials \textbf{#2}, #3 (#4)}
\newcommand{\NatMat}[4]{\textit{#1}, Nat. Mater. \textbf{#2}, #3 (#4)}
\newcommand{\AdvMater}[4]{\textit{#1}, Adv. Mater. \textbf{#2}, #3 (#4)}
\newcommand{\npjSpintron}[4]{\textit{#1}, npj Spintron. \textbf{#2}, #3 (#4)}
\newcommand{\APLQuantum}[4]{\textit{#1}, APL Quantum \textbf{#2}, #3 (#4)}
\newcommand{\NatMatonline}[4]{\textit{#1}, Nat. Mater. (#2)}

\begin{document}
\title{Hierarchical coupling in tripartite van der Waals cavity magnonics \\ for a low-loss and tunable coupled-mode Hamiltonian}

\author{Taehee Kim\protect\textsuperscript{*}}
\affiliation{DGIST Research Institute, DGIST, Daegu 42988, South Korea}

\author{Sunghun Park\protect\textsuperscript{*}}
\affiliation{Center for Theoretical Physics of Complex Systems, Institute for Basic Science (IBS), Daejeon 34126, South Korea }

\author{Byoung-moo Ann\protect\textsuperscript{*, \dag}}
\affiliation{Quantum Technology Institute, Korea Research Institute of Standards and Science, 34113 Daejeon, South Korea}

\author{Sung Jin An}
\affiliation{DGIST Research Institute, DGIST, Daegu 42988, South Korea}
\affiliation{Department of Physics and Chemistry, DGIST, Daegu 42988, Korea}

\author{Seung-Bo Shim}
\affiliation{Quantum Technology Institute, Korea Research Institute of Standards and Science, 34113 Daejeon, South Korea}

\author{Hakseong Kim}
\affiliation{Quantum Technology Institute, Korea Research Institute of Standards and Science, 34113 Daejeon, South Korea}

\author{Haydar Sahin}
\affiliation{Electrical and Computer Engineering, National University of Singapore, Singapore 117576, Republic of Singapore}

\author{Zhuo Bin Siu}
\affiliation{Electrical and Computer Engineering, National University of Singapore, Singapore 117576, Republic of Singapore}

\author{Mansoor B. A. Jalil}
\affiliation{Electrical and Computer Engineering, National University of Singapore, Singapore 117576, Republic of Singapore}

\author{Youngwook Kim}
\affiliation{Department of Physics and Chemistry, DGIST, Daegu 42988, Korea}

\author{Thomas K. M. Graham}
\affiliation{Department of Physics, Boston College, Chestnut Hill, Massachusetts 02467, USA}

\author{Brian B. Zhou}
\affiliation{Department of Physics, Boston College, Chestnut Hill, Massachusetts 02467, USA}

\author{Myung-Joong Hwang}
\affiliation{Division of Natural and Applied Sciences, Duke Kunshan University, Kunshan, Jiangsu 215300, China}

\author{Yaroslav M. Blanter}
\affiliation{Kavli Institute of Nanoscience, Delft University of Technology, Lorentzweg 1 2628CJ Delft, the Netherlands}

\author{Jungpil Seo}
\affiliation{Department of Physics and Chemistry, DGIST, Daegu 42988, Korea}

\author{Hee Chul Park\protect\textsuperscript{\ddag}}
\affiliation{Department of Physics, Pukyong National University, Busan 48513, South Korea}

\author{Minkyung Jung\protect\textsuperscript{\S}}
\affiliation{DGIST Research Institute, DGIST, Daegu 42988, South Korea}
\affiliation{Institute of next generation semiconductor technology (INST), DGIST, Daegu 42988, South Korea}
\affiliation{Department of Interdisciplinary Engineering, DGIST, Daegu 42988, South Korea}
\date{5th Nov 2026}

\begin{abstract}
One of the greatest merits of cavity magnonics (CM) is the ability to explore rich coupled-mode physics, driven by the large \textit{in situ} flux tunability of magnonic resonances and magnon--photon coupling.
Meanwhile, van der Waals (vdW) magnets offer an ultra-compact and highly integrated implementation of CM without complex fabrication processes.
Nonetheless, the substantial magnon losses significantly hinder the combination of all these advantages.
Here, we realize a hierarchical anti-crossing structure in a tripartite vdW CM with $\text{CrCl}_3$ and realize an effective coupled-dual-mode Hamiltonian.
We achieve an order-of-magnitude reduction in loss while maintaining a strong coupling condition. Additionally, we theoretically confirm the feasibility of simultaneous and \textit{in situ} tuning of both loss and coupling by varying the magnetic field direction and strength.
Featuring a micrometer-scale thickness and a width of a few millimeters, our device is significantly smaller than those based on bulk crystals or crystal arrays.
Our results establish vdW cavity magnonics as versatile platform for investigating intriguing phenomena predicted in coupled-mode systems.
\end{abstract}

\maketitle

\begingroup
\renewcommand{\thefootnote}{\protect\ensuremath{*}}
\footnotetext{These authors contributed equally to this work.}
\renewcommand{\thefootnote}{\protect\ensuremath{\dag}}
\footnotetext{byoungmoo.ann@kriss.re.kr}
\renewcommand{\thefootnote}{\protect\ensuremath{\ddag}}
\footnotetext{hc2725@gmail.com}
\renewcommand{\thefootnote}{\protect\ensuremath{\S}}
\footnotetext{minkyung.jung@dgist.ac.kr}
\endgroup

\section{Introduction}

Cavity magnonics (CM) has emerged as a powerful paradigm for enabling the efficient probing of magnon dynamics, coherent information processing, and the development of quantum hybrid systems \cite{Huebl-PRL-13, Zhang-PRL-14, Hisatomi-PRB-16, Harder-JAP-21, Rameshti-PhysRep-22, Tabuchi-PRL-14,Tabuchi-Science-15,Lachance-Science-20,Yuan-PhysRep-22,Fruy-NatMat-26, Xu-PRL-23, Rani-PRApplied-25, Kounalakis-PRL-22, Kounalakis-PRB-23}. 
A defining merit of CM lies in its capacity to explore rich coupled-mode physics and engineer microwave photonic modes, which is fundamentally driven by the large \textit{in situ} flux tunability of magnonic resonances and the magnon--photon coupling strengths \cite{Boventer-PRR-20}. 
The associated phenomena involve coherent perfect absorption (CPA) \cite{Rao-NatCommun-21, Shen-NatCommun-25}, exceptional points (EPs) \cite{Zhang-NatCommun-17, Zhang-PRL-19, Wang-PRL-25, Lambert-NatPhys-25}, topological effects \cite{Qian-PRL-24, Sun-APLQuantum-26}, ultrastrong coupling \cite{Ghirri-PRApplied-23, Guo-NanoLett-23} and other various phenomena \cite{Harder-PRL-18, Zhang-PRB-21, Zhang-npjSpintron-25, Bourcin-APLQuantum-26}.
The applications in these directions, however, require sufficiently low magnon damping, and consequently, most experimental studies implementing such intriguing coupled-mode physics with magnons have mainly employed low-damping systems such as bulk Yttrium iron garnet (YIG) crystals \cite{Zhang-NatCommun-17, Harder-PRL-18, Rao-NatCommun-21, Qian-PRL-24, Shen-NatCommun-25, Lambert-NatPhys-25, Sun-APLQuantum-26, Bourcin-APLQuantum-26}.

Meanwhile, van der Waals (vdW) materials have attracted immense attention for extending these capabilities toward the two-dimensional limit \cite{Gong-Nature-17, Huang-Nature-17, McGuire-PRMaterials-17, Burch-Nature-18, Zhang-NatComm-23, Park-RMP-26}. 
They offer a promising pathway for planar, ultra-compact and highly integrated cavity magnonic architectures while minimizing fabrication efforts \cite{Mandal-APL-20, Zhang-APL-21, Zollitsch-NatComm-23, Lee-CommPhys-23, Tang-NanoLett-25, Mandal-PRB-25, Almojuela-arXiv-26}.
Furthermore, vdW magnets provide unprecedented merits such as gate- and strain-induced tunabilities of their magnetic properties \cite{Deng-Nature-18, Cenker-NatNanotech-22}, and the ability to form suspended drumhead structures \cite{Fei-NanoLett-24}, alongside stronger nonlinear effects due to their small spin numbers \cite{Chen-NatCommun-21, Datta-NatMater-24}.

Despite the successful demonstration of strong cavity--magnon coupling in CM based on vdW magnets \cite{Mandal-APL-20, Zhang-APL-21, Tang-NanoLett-25, Mandal-PRB-25, Almojuela-arXiv-26}, substantial magnon dissipation remains a critical bottleneck.
This causes the signal to be obscured by the background near the cavity--magnon resonance, making it difficult to properly identify. Furthermore, this large dissipation severely complicates time-domain pulse-based measurements. Because the measurements must be completed within the extremely short timescale, strict bandwidth requirements are imposed on the measurement electronics. This tight time window also exacerbates pulse signal distortions and reflections caused by impedance mismatches at the input and output. Ultimately, these combined challenges substantially limit the utility of vdW cavity magnonics as a practical platform for exploring coupled-mode physics.

In this work, we devise and demonstrate a scheme to merge the advantages of vdW magnets and effectively realize a low-loss coupled-dual-mode Hamiltonian.
We mitigate the substantial magnon loss utilizing a tripartite coupling scheme, incorporating an auxiliary cavity mode into a conventional vdW cavity CM architecture. 
As experimental platform, we employ $\text{CrCl}_3$ and Nb superconducting planar resonators.
By optimizing the device parameters, we implement a hierarchical anti-crossing structure in the resonator transmission. While such structures have been reported in other systems~\cite{Long-PRL-18, Wright-JCP-23}, they remain unprecedented in CM. 
Unlike the sequential anti-crossings previously observed in multi-mode CM~\cite{Hyde-APL-16, Boventer-PRR-20, Kostylev-APL-16, Zhang-JAP-16}, this secondary anti-crossing sits on one of the primary branches. This exhibits a distinct dispersive-to-resonant transition and our analysis reveals that the damping of vdW magnons is reduced by an order of magnitude, manifesting the formation of strongly coupled and loss-mitigated two-mode Hamiltonian.

Our device features a micrometer-scale thickness and a width of a few millimeters, making it significantly smaller than systems based on low-damping bulk crystals or crystal arrays \cite{Zhang-NatCommun-17, Harder-PRL-18, Zhang-PRB-21, Rao-NatCommun-21, Qian-PRL-24, Shen-NatCommun-25, Lambert-NatPhys-25, Zhang-npjSpintron-25, Sun-APLQuantum-26, Bourcin-APLQuantum-26}. 
Although we investigate only a specific regime here, our theoretical model supports the feasibility of simultaneously tuning the coupling and dissipation. This capability will enable the exploration of various non-Hermitian phenomena using an ultra-compact platform. 
Additionally, we aim to theoretically investigate the implementation of this scheme with superconducting qubits.

\section{Results and discussion}

\subsection{Theoretical description.}

\begin{figure*}[htbp]
    \centering
    \includegraphics[width=0.9\linewidth]{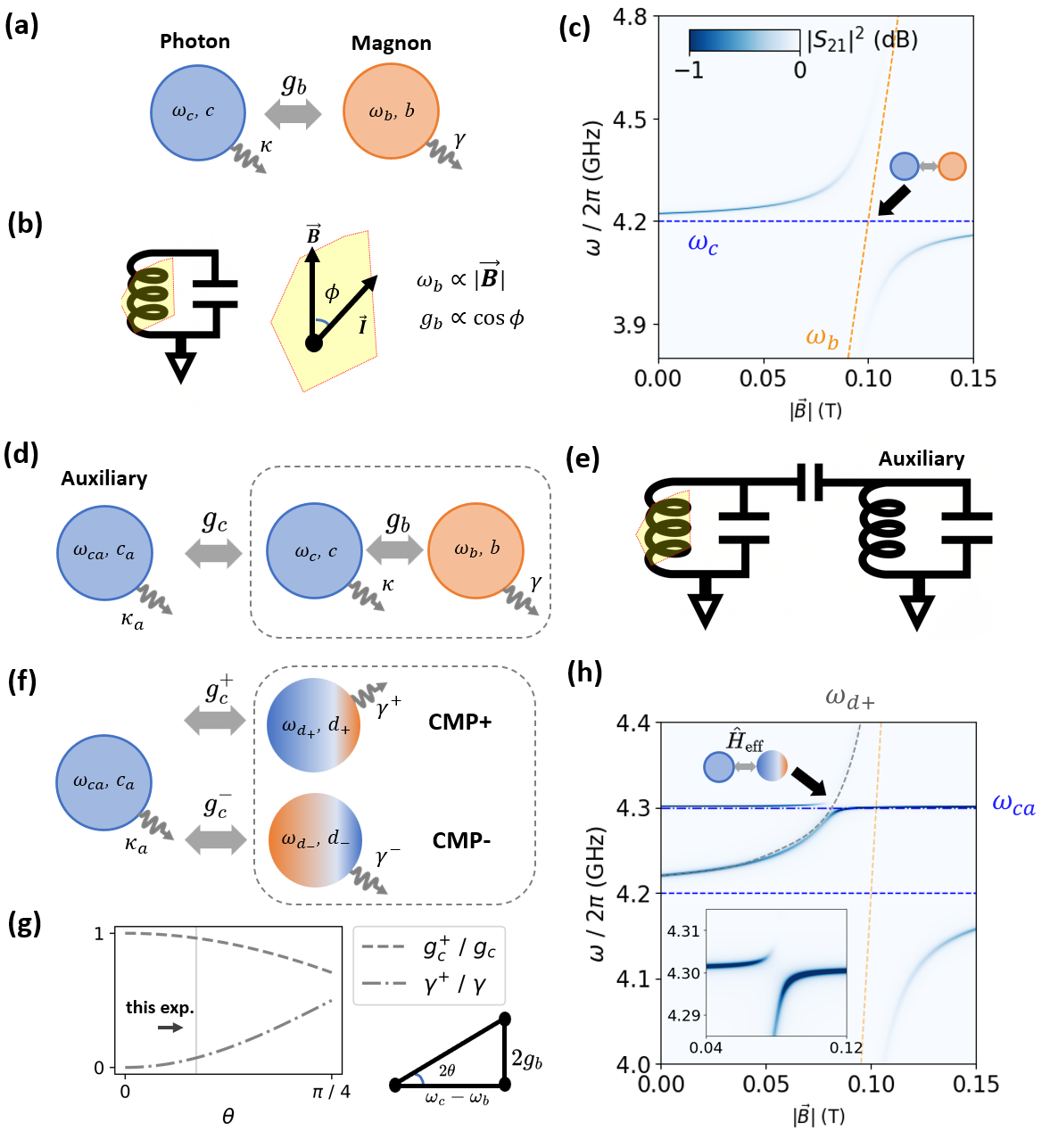}
    \caption{\textbf{Effective coupled-dual-mode Hamiltonian from tripartite hierarchical coupling with van der Waals cavity magnonics.}
    \textbf{(a)} Cavity magnonics (CM) configuration with resonant frequencies ($\omega_c, \omega_b$), field operators ($c, b$), dissipation rates ($\kappa, \gamma$), and coupling ($g_b$).
    \textbf{(b)} Circuit diagram. Yellow represents vdW flakes. $\vec{B}$ and $\vec{I}$ indicates in-plane magnetic field and current vector of the resonator. $\omega_b$ and $g_b$ are magnetically tunable as changing $|\vec{B}|$ and $\phi$.
    \textbf{(c)} Calculated $|S_{21}|^2$ of the cavity transmission with realistic parameters.
    \textbf{(d--e)} Tripartite coupling scheme, introducing an auxiliary cavity distinguished by subscripts $a$.
    \textbf{(f)} Formation of cavity magnon-polaritons (CMPs, dashed box). The gradations represent the contributions from the magnons and cavity fields to each CMP mode. The annihilation operators of the upper and lower CMPs (CMP$\pm$, respectively) are defined as $d_{\pm}$. Each mode possesses resonance frequencies $\omega_{d\pm}$ and damping rates $\gamma_{\pm}$, and interacts with the auxiliary cavity with coupling strengths $g_c^{\pm}$.
    \textbf{(g)} Calculated $g_c^+/g_c$ and $\gamma^+/\gamma$ with respect to the mixing angle $\theta$ between the cavity photons and magnons ($\theta = \frac{1}{2}\arctan\left(\frac{2g_b}{\omega_{c}-\omega_b}\right)$) when $\kappa=0$. One can \textit{in situ} tune $\theta$ as changing $|\vec{B}|$ and $\phi$, as presented in (b).
    We are focused on achieving a strong and resonant coupling between the CMP and auxiliary cavity mode ($\gamma^+, \kappa_a \ll g_c^+$, $\omega_{d_+}\approx\omega_{ca}$) with decent $g_c$ and sufficiently low $\theta$. A vertical line indicates the experimental regime.
    \textbf{(h)} Calculated $|S_{21}|^2$ in the tripartite configuration. Hierarchical anti-crossing structure appears, manifesting the realization of $\hat{H}_{\text{eff}}$. See the main text for the calculation parameters.}
    \label{fig:device_description}
\end{figure*}

Fig.~1(a,b) illustrates the configuration of the vdW cavity magnonic systems. Here, the resonance frequencies ($\omega_c, \omega_b$), annihilation operators ($c, b$), and dissipation rates ($\kappa, \gamma$) describe the cavity and magnon modes, respectively, with $g_b$ denoting the coupling strength between them. The yellow regions in the circuit diagrams represent the vdW flakes, while $\vec{B}$ and $\vec{I}$ denote the applied external magnetic field and the resonator current, respectively. As shown in the adjacent panels, $\omega_b$ and $g_b$ can be tuned by varying the magnetic field strength $|\vec{B}|$ and its orientation angle $\phi$.

We numerically calculate the resonator transmission $|S_{21}|^2$ using realistic experimental parameters. For the cavity, we set $\kappa/2\pi = 2.5\,\text{MHz}$, $\kappa_{\mathrm{ex}}/2\pi = 0.25\,\text{MHz}$, and $\omega_c/2\pi = 4.2\,\text{GHz}$. For magnons in Cr-based vdW materials, $\gamma/2\pi$ is on the order of $100\,\text{MHz}$, based on both theoretical predictions~\cite{Esteras-NanoLett-22} and experimental demonstrations~\cite{Zhang-APL-21}. In principle, both $\omega_c$ and $\kappa$ depend on $|\vec{B}|$, however, we omit this dependence in the calculations in Fig.~1 for simplicity, while explicitly accounting for it when analyzing the experimental data. For $\text{CrCl}_3$, the material chosen for our study, the magnon frequency $\omega_b$ exhibits an approximately linear dependence on $|\vec{B}|$ near $\omega_c$, with $|\vec{B}| \sim 100\,\text{mT}$ representing the typical field scale where $\omega_b \sim \omega_c$. Based on these considerations, we set $\gamma/2\pi = 100\,\text{MHz}$ and $\omega_b/2\pi = (42\,\text{GHz}\,\text{T}^{-1}) |\vec{B}|$.
We also set $g_b/2\pi=300\,\text{MHz}$, feasible value with sufficiently thick $\text{CrCl}_3$ flakes \cite{Mandal-APL-20, Zhang-APL-21}.

Fig.~1(c) shows the calculation result based on those parameters.
Although a signature of strong coupling emerges, it becomes faint near resonance owing to the large magnon dissipation rate. This substantially limits the utility of vdW cavity magnonics as a platform for exploring coupled-mode physics.

Fig. 1(d--e) describe the tripartite configuration. Here, we introduce an auxiliary cavity, distinguished by subscript $a$, is coupled with a strength $g_c$.
The system Hamiltonian reads
\begin{equation*}
\begin{split}
    \frac{H}{\hbar} =&  \left(\omega_{c} - i\kappa\right) c^\dagger c + \left(\omega_{ca} - i\kappa_a \right) c_a^\dagger c_a \\
    &+ \left(\omega_{b} - i\gamma \right) b^\dagger b \\
    &+ g_c (c^\dagger c_a + c_a^\dagger c) + g_b (b^\dagger c + c^\dagger b).
\end{split}
\end{equation*}
Here, $g_c$ indicates the coupling between the cavities.

In Fig. 1(f), we describe the system on the  cavity magnon--polaritons (CMPs) basis. We introduce $d_{\pm}$, the annihilation operators of the CMP as below
\begin{align}
    d_+ &= c \cos\theta + b \sin\theta, \label{eq:1} \\
    d_- &= c \sin\theta - b \cos\theta. \label{eq:2}
\end{align}
Here, $\theta = \frac{1}{2}\arctan\left(\frac{2g_b}{\omega_{c} - \omega_b}\right)$ ($0 \le \theta \le \frac{\pi}{4}$) is a mixing angle between the cavity photons and magnons. Confining ourselves to $\theta < \pi/4$, each CMP mode can be classified into upper ($+$) and lower ($-$) components. The damping rates $\gamma^{\pm}$ and the resonance frequencies $\omega_{d\pm}$ of each CMP mode are respectively given as below
\begin{align}
    \gamma^+ &= \kappa \cos^2\theta + \gamma \sin^2\theta, \label{eq:3} \\
    \gamma^- &= \kappa \sin^2\theta + \gamma \cos^2\theta, \label{eq:4}
\end{align}
and
\begin{equation}
    \omega_{d\pm} = \frac{\omega_{c} + \omega_b}{2} \pm \sqrt{(g_b)^2 + \left(\frac{\omega_{c} - \omega_b}{2}\right)^2}. \label{eq:5}
\end{equation}
Meanwhile, $g_c^+ = g_c \cos\theta$ and $g_c^- = g_c \sin\theta$ are satisfied. 
Then, we can recast the system Hamiltonian as below
\begin{equation}
\begin{split}
    \frac{H}{\hbar} =& \left(\omega_{ca} - i\kappa\right) c_a^\dagger c_a + \left(\omega_{d+} - i\gamma^+\right) d_+^\dagger d_+ \\
    &+ \left(\omega_{d-} - i\gamma_-\right) d_-^\dagger d_- \\
    &+ g_c^+ (d_+^\dagger c_a + c_a^\dagger d_+) + g_c^- (d_-^\dagger c_a + c_a^\dagger d_-).
\end{split}
\label{eq:6}
\end{equation}
Here, $\omega_{d+} \approx \omega_{c} + g_b \tan\theta$ and $g_b/2\pi$ is hundreds MHz large. 

For sufficiently small $g_c$, we achieve $|\omega_{ca} - \omega_{d-}| \gg g_c^-$ when $\omega_{d+}\approx\omega_{ca}$, and therefore, neglect the effects of the lower CMP.
We find that $g_c/2\pi \leq 20$ MHz is enough for this approximation.
Subsequently, the Hamiltonian is reduced to
\begin{equation}
\begin{split}
    \frac{H_{\text{eff}}}{\hbar} =& \left(\omega_{ca}  - i\kappa_a\right) c_a^\dagger c_a \\
    &+ \left(\omega_{d+} - i\gamma^+\right) d_+^\dagger d_+ + g_c^+ (d_+^\dagger c_a + c_a^\dagger d_+).
\end{split}
\label{eq:7}
\end{equation}

Fig. 1(g) shows the theoretical ratios of $g_{c}^+/g_c$ and $\gamma^{+}/\gamma$ with respect to $\theta$.
$\gamma^{+}$ can be approximately reduced by a factor of $\sin^2\theta$ when $\kappa$ is small whereas $g_{c}^+/g_c$ scales linearly with $\cos\theta$.
This result shows that strong and resonant coupling between the upper CMP and auxiliary cavity modes ($\gamma^+, \kappa_a \ll g_c^+$, $\omega_{d_+}\approx\omega_{ca}$) can be achieved properly selecting $\theta$ and $g_c$.
A gray vertical line indicates our experimental regime.
Although we focus on the regime of $g_c^{+}>\gamma^{+}$ and $\gamma^{+}/\gamma \ll 1$ in this study, one can also \textit{in situ} enter intermediate and weak coupling regimes by tuning $\theta$ and $\omega_b$.

Fig.~1(h) shows the numerically calculated transmission spectrum $|S_{21}|^2$ of the tripartite system.
Here, we set $\omega_{ca}/2\pi = 4.3\,\text{MHz}$, $\kappa_a/2\pi = 0.8\,\text{MHz}$, and $g_c/2\pi = 10\,\text{MHz}$, while the remaining parameters are kept identical to those in Fig.~1(c).
The result reveals a hierarchical coupling structure, where an secondary anti-crossing, corresponding with $\hat{H}_{\text{eff}}$[Eq.~(7)], formed along one of the primary branches.
The absence of the upper branch in the secondary anti-crossing is attributed to the formation of a dark mode, arising from the assumption that both cavities are coupled to the same feedline. This characteristic feature is also observed in the experiments discussed later.
This demonstrates the feasibility of substantially enhancing system's utility as an effective coupled-mode Hamiltonian.
This exhibits a distinct dispersive-to-resonant transition whereas the primary anti-crossing becomes almost faint at the resonance.

See \textbf{Supplementary Materials} for the details on the all theoretical derivations and numerical calculations.

\subsection{Cavity magnonics with $\text{CrCl}_3$}

\begin{figure}[htbp]
    \centering
    \includegraphics[width=1.0\linewidth]{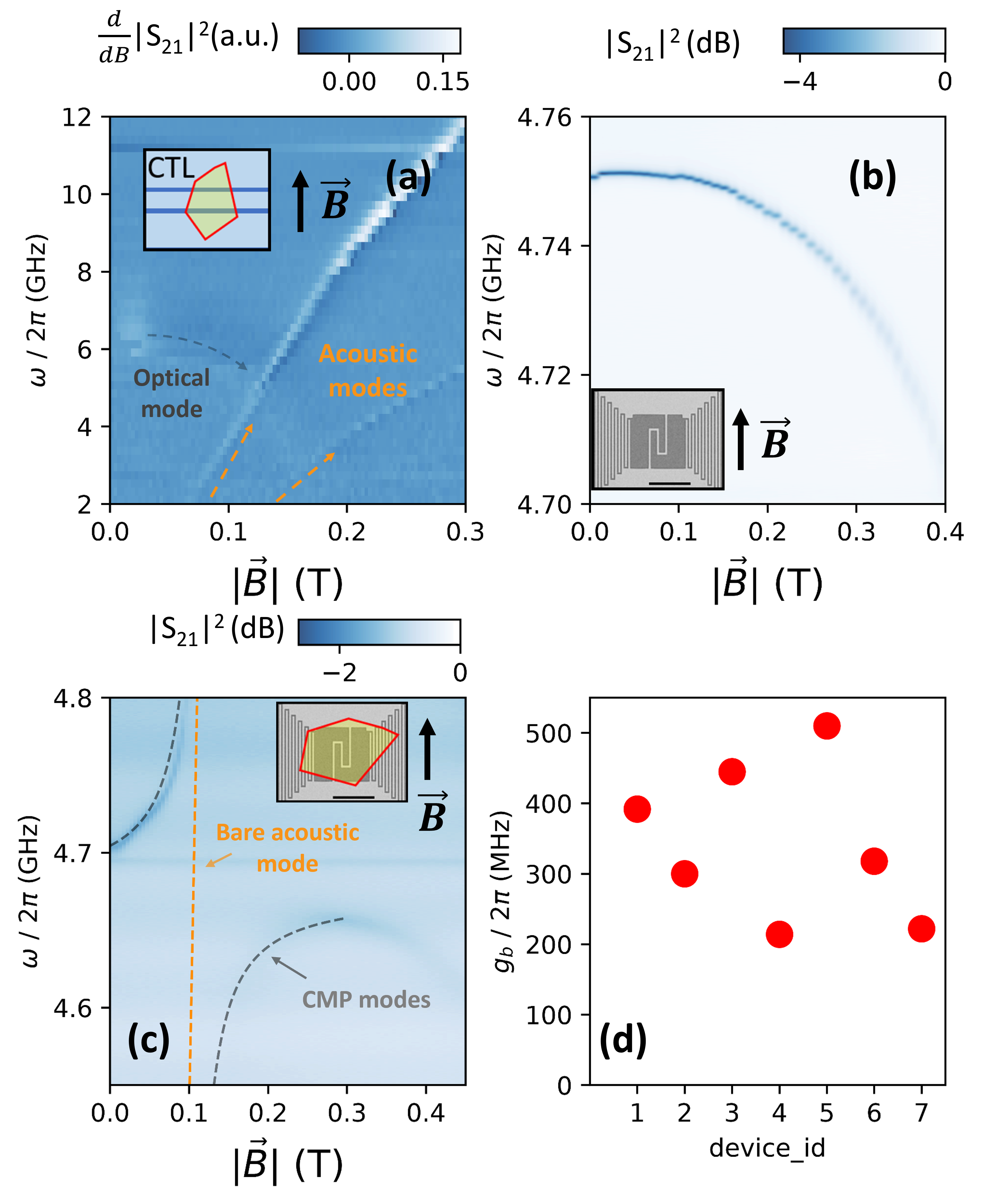}
    \caption{\textbf{Characterizing cavity and magnon modes.} 
    \textbf{(a)} A $\text{CrCl}_3$ flake is placed on a copper transmission line (inset). The differential magnitude of transmission ($\frac{d}{dB}|S_{21}|^2$) with respect to probe frequency $\omega/2\pi$ and in-plane magnetic field $|\vec{B}|$ is presented, which shows the two acoustic magnon modes (orange arrows) and an optical magnon mode (dark arrow) of the $\text{CrCl}_3$ flake. The data is taken at $T = 3$~K. 
    \textbf{(b)} Transmission $|S_{21}|^2$ of a bare superconducting niobium lumped-element resonator (LER) measured at $T = 1.7$~K. The LER is capacitively side-coupled to the feedline. The resonance decreases with increasing in-plane magnetic field. The inset shows the inductor and finger capacitor part of the LER.
    \textbf{(c)} $|S_{21}|^2$ of an LER with a $\text{CrCl}_3$ flake. Strong frequency splitting at $|\vec{B}| = 0.1$~T, where the resonator frequency matches the magnetic resonance frequency, indicates strong coupling between the resonator and the magnons of $\text{CrCl}_3$. Black dashed curves represent fits to the data, while yellow dashed lines indicate the bare resonator and magnon modes ($\omega_b/2\pi = 42 \times |\vec{B}| $ GHz/T). We attribute the horizontal line around $\omega/2\pi = 4.69$~GHz to an unknown stray mode of the system.
    \textbf{(d)} The cavity--magnon coupling strengths $g_b/2\pi$ extracted from seven measured devices. All scale bars in the insets represent 5~$\mu$m.}
    \label{fig:characterization}
\end{figure}

In this section, we present characterizing experiments of vdW CM with a single microwave mode as a preliminarily step before introducing the tripartite configuration.
The sample is mounted in a cryostat with temperature under $T=3 K$.
Strong coupling ($g_b > \gamma$) is achievable using sufficiently thick vdW flakes \cite{Mandal-APL-20, Zhang-APL-21} or high-spin density materials coupled to low-impedance resonators \cite{Almojuela-arXiv-26}. Here, we use flakes ($\sim 20\,\mu\text{m}$) of $\text{CrCl}_3$ and lumped-element resonators (LERs) defined on a planar niobium films. This configuration provides us with sufficient cavity--magnon coupling strengths to implement the tripartite configuration. 

In Figs.~2(a) and 2(b), we characterize the antiferromagnetic resonance of the $\text{CrCl}_3$ and the resonators. A $\text{CrCl}_3$ flake is positioned on a coplanar waveguide and secured by placing Kapton tape directly over it (inset of Fig.~2(a)).
We probe the differential transmission spectra ($\frac{d}{dB}|S_{21}|^2$) while scanning the in-plane magnetic field $|\vec{B}|$ (Fig. 2(a)). We identify acoustic and optical magnon modes from the data. These modes in $\text{CrCl}_3$ flakes have also been reported in previous studies \cite{MacNeill-PRL-19, Wang-NanoLett-19, Kapoor-AdvMater-21}. In this work, we consider only the acoustic mode marked by the left orange arrow. Fig. 2(b) shows the transmission $|S_{21}|^2$ of a superconducting cavity made by a LER defined on a 100 nm thick niobium film on SiO substrate. The observed downward shift in the resonance with increasing $|\vec{B}|$ indicates the enhanced kinetic inductance of the superconducting wire, attributable to the Cooper-pair-breaking effect induced by the magnetic field.

The coupling between the microwave modes and magnons of $\text{CrCl}_3$ are investigated in Fig. 2(c--d). A single crystal flake of $\text{CrCl}_3$ is carefully positioned on the cavities. 
Subsequently, we investigate the interaction between the magnetic resonance of $\text{CrCl}_3$ and the superconducting cavity. Fig. 2(c) clearly shows the strong coupling between the magnon and cavity modes. We fit the frequency splitting using Eq. (5). The extracted coupling rate $g_b / 2\pi$ in this sample is estimated to be $302\text{ MHz}$. We repeat the experiments on seven distinct samples and present the extracted coupling rates in Fig. 2(d). $g_b / 2\pi$ varies from sample to sample, likely due to differences in the distance between the resonator and the $\text{CrCl}_3$ flake. The highest observed $g_b / 2\pi$ in our experiments reaches $510\text{ MHz}$.

\subsection{Device for tripartite coupling}

\begin{figure}[htbp]
    \centering
    \includegraphics[width=1.0\linewidth]{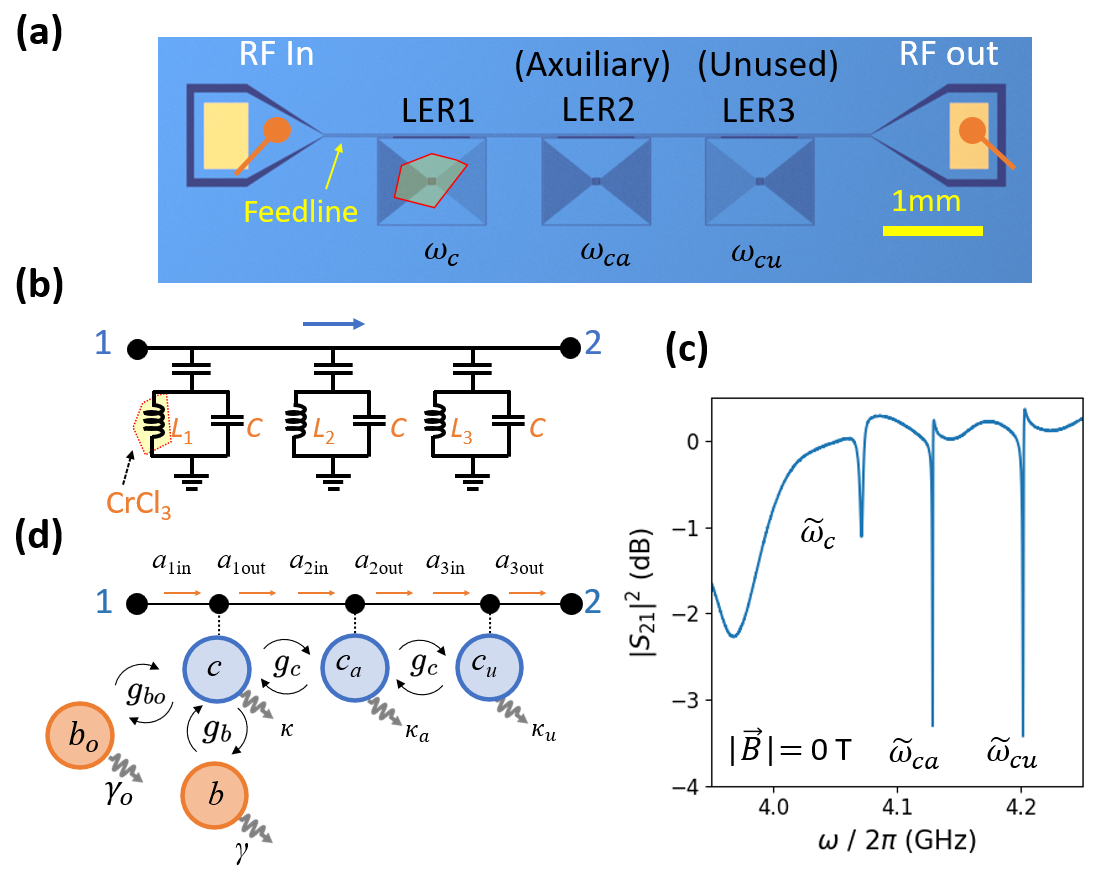}
    \caption{\textbf{Device description.} 
    \textbf{(a)} Schematic of the device. An array of three superconducting lumped-element resonators (LERs) is coupled to a feedline. Their bare resonant frequencies are presented below ($\omega_{c}$, $\omega_{ca}$, $\omega_{cu}$). LER1 and LER2 serve as auxiliary and target cavity modes in Fig. 1(b). We use LER3 only for reference. A $\text{CrCl}_3$ flake (yellow) sits on the inductor part of LER1.
    \textbf{(b)} Equivalent circuit diagram. The LERs have different inductances $L_1$, $L_2$, and $L_3$ with the same capacitance $C$.
    \textbf{(c)} The observed transmission spectrum $|S_{21}|^2$ of the device in (a) as a function of probe frequency. The device is cooled down to $T = 1.7$~K and $|\vec{B}| = 0$~T. The dips correspond to the eigenfrequencies $\tilde{\omega}_{c}$, $\tilde{\omega}_{ca}$, $\tilde{\omega}_{cu}$ of $H_{\text{dev}}$ in $|\vec{B}| = 0$~T, which are adiabatically connected to $\omega_{c}$, $\omega_{ca}$, and $\omega_{cu}$ as $g_c \rightarrow 0$. We extract $\kappa/2\pi \approx 3.0\text{ MHz}$, $\kappa_{a,u}/2\pi \approx 0.8\text{ MHz}$ from this data. We attribute the large difference between $\kappa_{a,u}$ and $\kappa$ to the flake-transfer contamination. $\kappa_{a,u}$ are similar to the dissipation rates of the cavities without flakes, which indicates that they are negligibly affected by the flake-induced contamination.
    \textbf{(d)} The diagram represents the modes and couplings among the modes. An optical magnon mode ($b_o$) coupled to the cavity is added to the diagram.}
    \label{fig:device_description}
\end{figure}

In Fig. 3(a), we present a schematic of the device to implement the tripartite coupling in Fig. 1. Multiple LERs are capacitively coupled to an RF feedline. Each LER is also coupled to their nearest neighbors. LER1 and LER2 respectively act as the cavity modes in Fig. 1(b). LER3 is not directly relevant to the focus of our study and used only as a reference. A $\text{CrCl}_3$ flake is positioned on the LER1 and secured with Kapton tape. 

Fig. 3(b) presents an equivalent circuit diagram of the feedline with the LERs. Each LER has different inductance ($L_1$, $L_2$, and $L_3$) and approximately the same capacitance. Fig. 3(c) displays the transmission spectra ($|S_{21}|^2$) of resonances as a function of frequency at $B = 0\text{ T}$. As predicted, three distinct resonance frequencies are observed within the frequency range of 4.0 to 4.2 GHz. From the linewidths and dips of the resonances, we extract $\kappa/2\pi \approx 3.0\text{ MHz}$, $\kappa_{a,u}/2\pi \approx 0.8\text{ MHz}$, and $\kappa_{ex}/2\pi\approx0.25\text{ MHz}$. 
Here, $\kappa_{ex}$ is the external coupling rate of the cavities to the feedline.  $\kappa_{a,u}$ of the auxiliary and unused cavities are similar to the dissipation rates of the cavities without flakes. The large difference between $\kappa_{a,u}$ and $\kappa_a$ is due to the contamination of the cavity while placing the flake.

The schematic diagram in Fig. 3(d) illustrates the coupling mechanism between the magnon and the resonators. The auxiliary mode is coupled to the magnon of $\text{CrCl}_3$ with a coupling strength $g_b$. The cavities are capacitively coupled with a coupling strength $g_c$. Here, $c_u$ and $b_o$ are the annihilation operators of the modes not directly relevant to our study (LER3 and optical magnon mode). $\omega_{cu}$ and $\omega_{bo}$ respectively indicate their bare resonant frequencies. $g_{bo}$ indicates the coupling strength between the auxiliary cavity and optical magnon modes. The LER3 is capacitively coupled with auxiliary cavity with a coupling strength $g_c$.

The device Hamiltonian is then given by
\begin{equation}
    H_{\text{dev}} = H + H_{cu} + H_{bo}.
    \label{eq:8}
\end{equation}
$H$ is the tripartite Hamiltonian introduced in Fig. 1(b). $H_{cu}/\hbar = (\omega_{cu} - i\kappa_u) c_u^\dagger c_u + g_c(c_a^\dagger c_u + c_u^\dagger c_a)$ and $H_{bo}/\hbar = (\omega_{bo} - i\gamma_o) b_o^\dagger b_o + g_{bo}(c^\dagger b_o + b_o^\dagger c)$ are the Hamiltonian contributed by the optical magnon and unused cavity modes, respectively.

\begin{figure*}[htbp]
    \centering
    \includegraphics[width=1.0\linewidth]{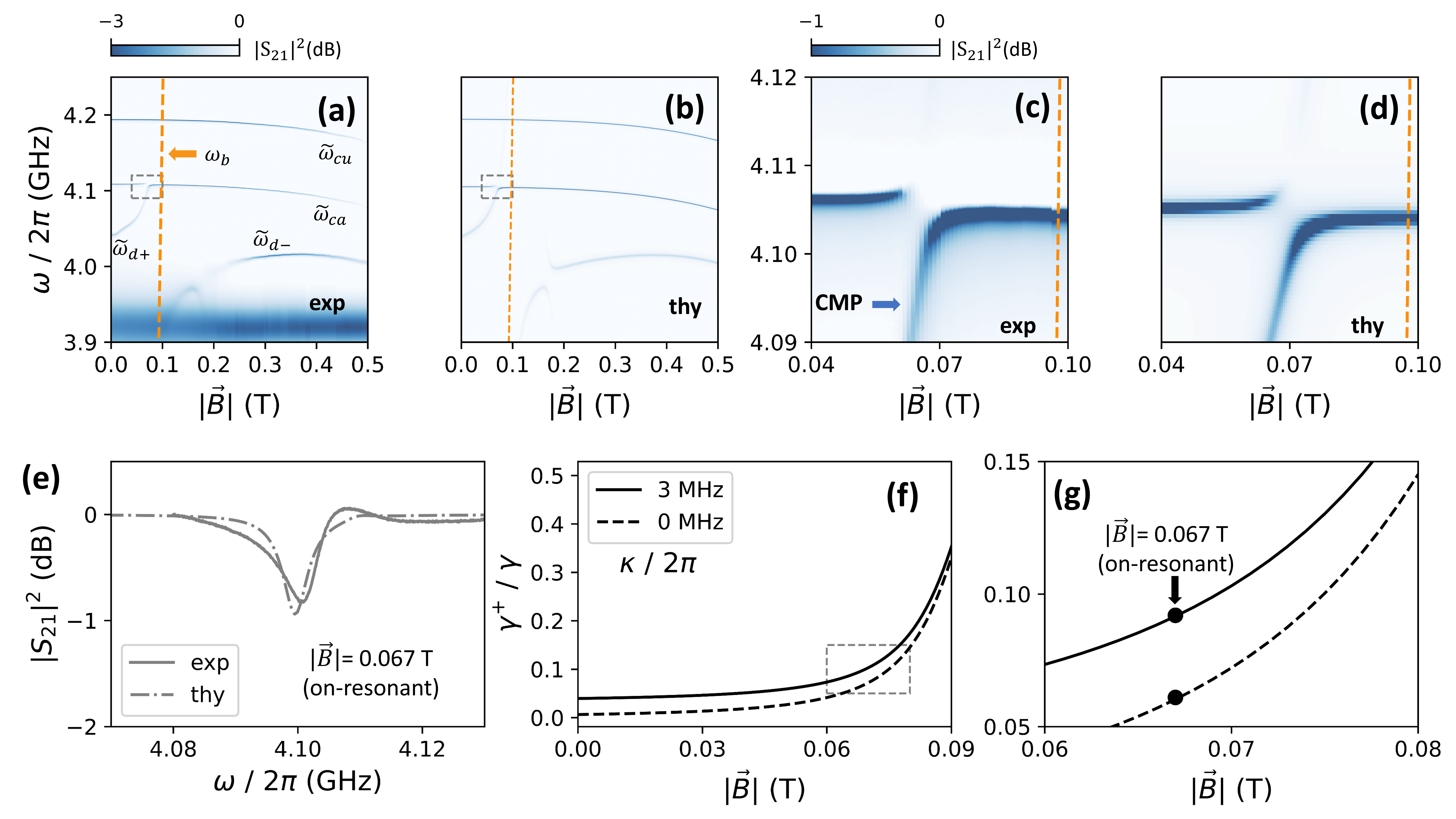}
    \caption{\textbf{Experimental realization of low-loss coupled-dual-mode Hamiltonian.} 
    \textbf{(a)} The observed transmission spectrum $|S_{21}|^2$ of the device in Fig. 3(a) as a function of probe frequency ($\omega/2\pi$) and magnitude of in-plane magnetic field ($|\vec{B}|$) at $T = 1.7$~K. $\tilde{\omega}_{d+}$, $\tilde{\omega}_{d-}$, $\tilde{\omega}_{ca}$ and $\tilde{\omega}_{cu}$ are eigenfrequencies of $H_{\text{dev}}$, adiabatically connected to $\omega_{d+}$, $\omega_{d-}$, $\omega_{ca}$ and $\omega_{cu}$ as $g_c \rightarrow 0$. 
    \textbf{(b)} Calculated transmission spectrum $|S_{21}|^2$, which exhibits excellent agreements with the observation. 
    \textbf{(c--d)} Magnification of the dashed boxes in (a) and (b).
    \textbf{(e)} Comparison of the calculated and experimental $|S_{21}|^2$ when the CMP and auxiliary cavity mode are resonant. The strong coupling is demonstrated between two modes, and the auxiliary cavity clearly shows its identity.
    \textbf{(f--g)} Estimated $\gamma^+/\gamma$ with respect to $|\vec{B}|$. (g) is the magnification of the dashed box in (f). The circles in (g--h) indicate the points when the CMP mode becomes resonant to the auxiliary cavity mode.}
    \label{fig:resonant_interaction}
\end{figure*}

\subsection{Experimental realization}

Fig. 4(a–b) shows the experimentally observed (a) and numerically calculated (b) transmission of the device presented in Fig. 3(a) with respect to in-plane magnetic field $|\vec{B}|$. Probe power in the experiment was $-50$ dBm at the device. $\tilde{\omega}_{d+}$, $\tilde{\omega}_{d-}$, $\tilde{\omega}_{ca}$ and $\tilde{\omega}_{cu}$ are eigenfrequencies of $H_{\text{dev}}$, adiabatically connected to $\omega_{d+}$, $\omega_{d-}$, $\omega_{ca}$ and $\omega_{cu}$ as $g_c \rightarrow 0$.

A clear hierarchical anti-crossing structure is observed in the resonator transmission. The secondary
anti-crossing (inside dashed box) sits on one of the primary branches.
The observation and calculation agree excellently with each other. From the numerical calculation, we extract $g_b/2\pi \approx 325\text{ MHz}$ and $\kappa/2\pi \approx 3.0\text{ MHz}$, $\kappa_{a,u}/2\pi \approx 0.8\text{ MHz}$, $g_c/2\pi \approx 5.5\text{ MHz}$ and $\gamma/2\pi \approx 90\text{ MHz}$ by comparing the observation and calculation. We use $\omega_b/2\pi = 42 \text{ GHz/T} \times |\vec{B}|$ in the calculation. See \textbf{Supplementary Materials} for the details on the numerical calculation.

Fig. 4(c--d) provide magnified views of the dashed boxes in Fig. 4(a--b). Fig. 4(e) shows a line-cut when the CMP and cavity modes are on-resonant. The clear anti-crossing represents the strong and resonant coupling between the upper CMP and auxiliary cavity. The upper branch of the anti-crossing is invisible as they form a dark state to the feedline.
From the anti-crossing data, the extracted coupling strength between them is $g_c^+/2\pi \approx 5.3\text{ MHz}$. The ratio  $g_c^+/g_c$ is consistent with the theoretical prediction.

The estimated $\gamma^+$ with respect to $|\vec{B}|$ is presented in Figs.~4(f) and 4(g) as solid lines. 
Around the resonance ($|\vec{B}| \approx 0.067\,\mathrm{T}$), $\gamma^+$ is an order of magnitude smaller than $\gamma$.
Assuming an ideal scenario where the intrinsic losses of the auxiliary cavity are negligible (dashed line), a lower $\gamma^+$ can be achieved for a given $|\vec{B}|$. 
Assuming an ideal situation ($\kappa = 0$), $\gamma^+$ can be further reduced.

We demonstrated a similar operation in another device featuring the same configuration as Fig.~3(a). See the \textbf{Supplementary Materials} for the experimental data from this additional device.

\subsection{Toward nonlinear coupled-mode Hamiltonian}

\begin{figure*}[htbp]
    \centering
    \includegraphics[width=1.0\linewidth]{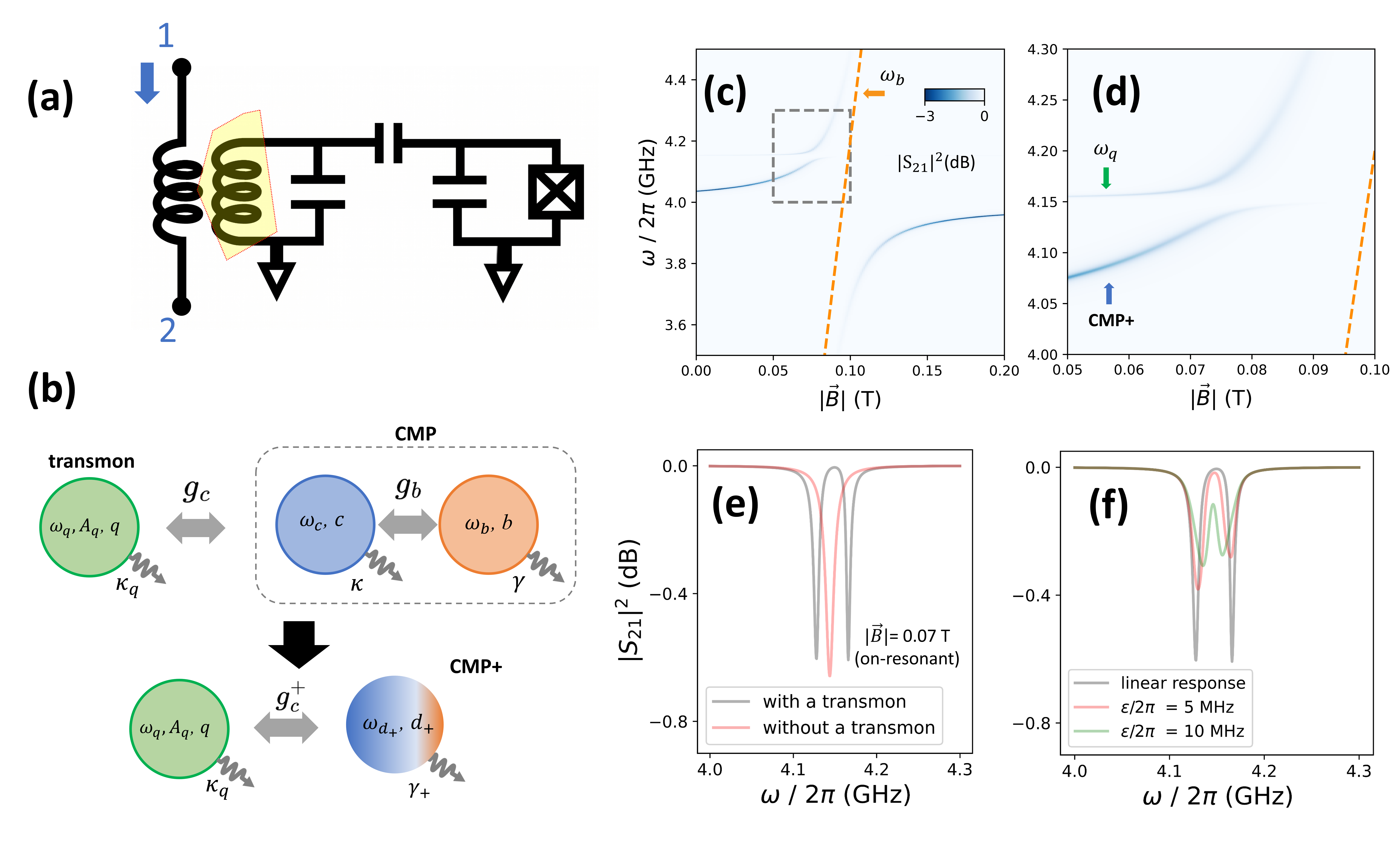}
    \caption{\textbf{Simulation of tripartite hierarchical  coupling with a vdW magnet, microwave resonator, and transmon.}
    \textbf{(a)} Equivalent circuit diagram of the microwave cavity inductively coupled to a feedline (ports 1 to 2) and capacitively connected to a transmon qubit. 
    \textbf{(b)} Schematic diagram of the system interactions. The bare transmon (frequency of $\omega_q$, anharmonicity of $A_q$, creation operator $q$) is coupled to the cavity mode with a coupling strength $g_c$, while the cavity is coupled to the magnon mode. As in Fig. 1, an effective model where the CMP is treated as a single upper CMP mode ($\omega_{d+}$) interacting with the transmon via an effective coupling $g_c^+$. In the numerical calculation, we use $\omega_q/2\pi=4.15$ GHz, $\kappa_q/2\pi=1$ MHz, $A_q/2\pi=200$ MHz, and $g_c/2\pi=20$ MHz. For the parameters on the magnons, auxiliary resonator, and their coupling and dissipations, we use the same values in Fig. 4. 
    \textbf{(c)} $|S_{21}|^2$ of resonator transmission as a function of $|\vec{B}|$. The orange dashed line represents the magnon frequency ($\omega_b$), showing a clear anti crossing with the cavity mode. 
    \textbf{(d)} Zoom-in view of the dashed rectangular region in (c), illustrating the interaction between the transmon ($\omega_q$) and the CMP mode.
    \textbf{(e)} $|S_{21}|^2$ at a fixed magnetic field of $B = 0.07$~T with (grey line) and without (pink line) the transmon qubit.
    \textbf{(f)} $|S_{21}|^2$  for various probe strengths $\epsilon/2\pi$. Compared to the linear response (grey line), increasing the probe strength to $5$~MHz (pink line) and $10$~MHz (green line) induces a noticeable deformation in the lineshape due to the saturation of the transmon qubit.}
    \label{fig:transmon_interface}
\end{figure*}

Incorporating highly nonlinear degrees of freedom like superconducting qubits (SCQs) to the systems significantly enhances coupled-mode dynamics.
Until now, theoretical and experimental applications of SCQs in CM \cite{Tabuchi-Science-15, Lachance-Science-20, Rani-PRApplied-25, Kounalakis-PRL-22, Kounalakis-PRB-23, Xu-PRL-23} have primarily involved Yttrium Iron Garnet (YIG) crystals, leaving vdW magnets largely unexplored because of the higher damping. 
Here, we numerically demonstrate that our scheme is also viable with SCQs.

Fig. 5(a) shows a schematic of the proposed system.
A superconducting transmon qubit \cite{Koch-PRA-2007} is capacitively coupled to a resonator. The resonator is probed by an inductively coupled feedline. At the same time, vdW flakes sit on the inductor part of the resonator.
Fig. 5(b) presents a schematic description of the system including interactions. The system is identical with Fig. 1(b) except that the auxiliary cavity is replaced with the transmon. The total Hamiltonian of the system is identical with Eq. (1) except that the target cavity Hamiltonian is replaced with that of the transmon, which reads
\begin{equation*}
    \frac{H_q}{\hbar} = (\omega_q - i\kappa_q ) q^\dagger q - \frac{A_q}{2} q^\dagger q^\dagger q q.
\end{equation*}
Here, $\omega_q$, $\kappa_q$, and $A_q$ are the frequency of fundamental transition, dissipation rate, and anharmonicity of the transmon.
In the numerical calculation, we use $\omega_q/2\pi=4.15$ GHz, $\kappa_q/2\pi=1$ MHz, $A_q/2\pi=200$ MHz, and $g_c/2\pi=20$ MHz.
These are realistic values in experiments. Particularly, $\kappa_q/2\pi=1$ MHz of transmons is surely achievable under in-plane magnetic field up to a few hundreds mT, as demonstrated in \cite{Krause-PRApp-22}.
For the parameters on the magnons, resonator, and their coupling and dissipation rates, we use the same values obtained in Fig. 4.

We numerically solve the Lindblad master equation of the total Hamiltonian including the dissipation. 
We solve the dynamics of the tripartite system and obtain the reduced density matrix by tracing out all but the resonator degrees of freedom.
We use the Quantum Toolbox in Python (QuTiP) version 5.1.1. \cite{Qutip1,Qutip2}.
Fig. 5(c) shows the numerically calculated transmission spectrum $|S_{21}|^2$ as a function of $|\vec{B}|$, revealing a distinct anti-crossing between the cavity mode and the magnon frequency ($\omega_b$, orange dashed line).
The probe strength ($\epsilon$) used in the calculation is small enough to show the linear response.
To investigate the interaction with the transmon qubit in detail, Fig. 5(d) provides a magnified view of the dashed rectangular region highlighted in Fig. 5(c).
This panel demonstrates the strong coherent coupling between the transmon ($\omega_q$) and the upper CMP mode with respect to $|\vec{B}|$.

$|S_{21}|^2$ in Fig. 5(e) highlights the explicit impact of the transmon qubit. 
In contrast to the case without the transmon (pink line), $|S_{21}|^2$ with the transmon (grey line) clearly exhibits hybridized features resulting from the coherent interaction between the transmon and CMP mode.
Fig. 5(f) illustrates the nonlinear transmission responses for various drive strengths $\epsilon/2\pi$. 
Compared to the linear response (grey line), increasing the drive strength to $5$~MHz (pink line) and $10$~MHz (green line) induces a significant deformation in the lineshape, which is directly attributed to the saturation of the transmon qubit.

See the \textbf{Supplementary Materials} for the details on the numerical simulation.

\section{Conclusion}

In conclusion, we have devised and experimentally demonstrated a novel tripartite hierarchical coupling architecture that effectively mitigates the critical bottlenecks of large damping rates of vdW magnets. 
While this study focuses on the strong coupling regime, our theoretical framework enables simultaneous tuning of coupling and dissipation, which counts to explore diverse non-Hermitian phenomena. 
We use a micrometer-thick $\text{CrCl}_3$ flake and superconducting niobium coplanar resonators for experimental realization, substantially more compact than conventional bulk crystal or array systems.

Our study establishes vdW cavity magnonics as a miniaturized, scalable, and versatile platform for investigating intriguing phenomena predicted in coupled-mode systems, including coherent perfect absorption, exceptional points, topological effects, and many others.
Our scheme can be extended to any other vdW materials other than $\text{CrCl}_3$. with the strong coupling to microwave cavities.
For instance, applying this scheme to high-spin-density vdW magnets such as $\mathrm{Cr}_2\mathrm{Ge}_2\mathrm{Te}_6$ (CGT) will enable the utilization of nanometer-thin flakes, opening the ability to probe and exploit exotic, unprecedented 2D magnetic phenomena at the atomic scale regime.
Ultimately, these results bridge the merits of vdW magnets and cavity magnonics toward ultra-compact and planar platform implementing coupled-mode physics.

\section{Acknowledgements}
We thank Jinuk Kim for fruitful discussions.
This work was supported by the National Research Foundation of Korea (NRF) grant funded by the Korea government (MSIT)(RS-2023-00213037) and Korea Research Institute of Standards and Science (KRISS-GP2026-0010-05). 
We also thanks for the support in device fabrication at the Nanofab, Korea Research Institute of Standards and Science.
S.P. acknowledges the support from the Institute for Basic Science (IBS) in the Republic of Korea through the project IBS-R024-Y4.
This work was also supported by the Mid-Career Researcher Program of the National Research Foundation of Korea (NRF) (RS-2023-NR076585), the DGIST R\&D Program of the Ministry of Science and ICT (26-ET-02), the Strategic Research Program under the DGIST R\&D Program of the Ministry of Science and ICT (26-SR-01, 26-SR-02, 26-SR-03) and NRF grants funded by the Korean government (MSIT \& MOE) (RS-2023-00278511, RS-2023-00256050, RS-2024-00402302, RS-2025-25463923, RS-2025-02315685). 
B.B.Z. and T.K.M.G. were supported by the National Science Foundation Award DMR-2047214.
Part of this work was supported by the Global Partnership Program of Leading Universities in Quantum Science and Technology (RS-2025-02317602). This work was also supported by the QuantERA ENTUNE project, funded by the NRF (No. RS-2026-25468343) and co-funded by the European Commission (No. 101212998). We also acknowledge the partner group program of the Max Planck Society.

Large language models (Gemini Pro 3.1) were used for language editing and proofreading of this manuscript. The authors have reviewed, edited, and approved the final text and take full responsibility for the content of the publication.

\section*{Author contributions}
\textbf{T.K} -- Data curation, Formal analysis, Investigation,  Resources, Writing – review \& editing,
\textbf{S.P} -- Data curation, Formal analysis, Investigation, Validation, Visualization, Writing – review \& editing
\textbf{B.A} -- Conceptualization, Data curation, Formal analysis, Funding acquisition, Investigation, Methodology, Project administration, Resources, Software, Supervision, Validation, Visualization, Writing – original draft, Writing – review \& editing,
\textbf{S.J.A} -- Investigation, Resources,
\textbf{S.S} -- Resources,
\textbf{H.K} -- Resources,
\textbf{H.S} -- Resources,
\textbf{Z.B.S} -- Resources,
\textbf{M.B.A.J} -- Resources,
\textbf{Y.K} -- Funding acquisition, Resources,
\textbf{T.K.M.G} -- Resources, Writing – review \& editing,
\textbf{B.B.Z} -- Resources, Writing – review \& editing,
\textbf{M.H} -- Investigation, Writing – review \& editing,
\textbf{Y.M.B} -- Investigation, Writing – review \& editing,
\textbf{J.S} -- Funding acquisition, Resources,
\textbf{H.C.P} -- Conceptualization, Data curation, Formal analysis, Funding acquisition, Investigation, Methodology, Project administration, Resources, Software, Supervision, Validation, Visualization, Writing – review \& editing,
\textbf{M.J} -- Conceptualization, Data curation, Formal analysis, Funding acquisition, Investigation, Methodology, Project administration, Resources, Software, Supervision, Validation, Visualization, Writing – review \& editing.

\section*{Competing interests}
The authors declare no competing interests.

\section*{Data availability}
Data supporting the plots within the main text of this paper is available from the corresponding author upon reasonable request.

\section*{Code availability}
Code used to produce the plots within this paper is available from the corresponding author upon reasonable request.

\end{document}

\title{Hierarchical coupling in tripartite van der Waals cavity magnonics \\ for a low-loss and tunable coupled-mode Hamiltonian}

\author{Taehee Kim\protect\textsuperscript{*}}
\affiliation{DGIST Research Institute, DGIST, Daegu 42988, South Korea}

\author{Sunghun Park\protect\textsuperscript{*}}
\affiliation{Center for Theoretical Physics of Complex Systems, Institute for Basic Science (IBS), Daejeon 34126, South Korea }

\author{Byoung-moo Ann\protect\textsuperscript{*,\,\dag,\,\P}}
\affiliation{Quantum Technology Institute, Korea Research Institute of Standards and Science, 34113 Daejeon, South Korea}

\author{Sung Jin An}
\affiliation{DGIST Research Institute, DGIST, Daegu 42988, South Korea}
\affiliation{Department of Physics and Chemistry, DGIST, Daegu 42988, Korea}

\author{Seung-Bo Shim}
\affiliation{Quantum Technology Institute, Korea Research Institute of Standards and Science, 34113 Daejeon, South Korea}

\author{Hakseong Kim}
\affiliation{Quantum Technology Institute, Korea Research Institute of Standards and Science, 34113 Daejeon, South Korea}

\author{Haydar Sahin}
\affiliation{Electrical and Computer Engineering, National University of Singapore, Singapore 117576, Republic of Singapore}

\author{Zhuo Bin Siu}
\affiliation{Electrical and Computer Engineering, National University of Singapore, Singapore 117576, Republic of Singapore}

\author{Mansoor B. A. Jalil}
\affiliation{Electrical and Computer Engineering, National University of Singapore, Singapore 117576, Republic of Singapore}

\author{Youngwook Kim}
\affiliation{Department of Physics and Chemistry, DGIST, Daegu 42988, Korea}

\author{Thomas K. M. Graham}
\affiliation{Department of Physics, Boston College, Chestnut Hill, Massachusetts 02467, USA}

\author{Brian B. Zhou}
\affiliation{Department of Physics, Boston College, Chestnut Hill, Massachusetts 02467, USA}

\author{Myung-Joong Hwang}
\affiliation{Division of Natural and Applied Sciences, Duke Kunshan University, Kunshan, Jiangsu 215300, China}

\author{Yaroslav M. Blanter}
\affiliation{Kavli Institute of Nanoscience, Delft University of Technology, Lorentzweg 1 2628CJ Delft, the Netherlands}

\author{Jungpil Seo}
\affiliation{Department of Physics and Chemistry, DGIST, Daegu 42988, Korea}

\author{Hee Chul Park\protect\textsuperscript{\ddag}}
\affiliation{Department of Physics, Pukyong National University, Busan 48513, South Korea}

\author{Minkyung Jung\protect\textsuperscript{\S}}
\affiliation{DGIST Research Institute, DGIST, Daegu 42988, South Korea}
\affiliation{Institute of next generation semiconductor technology (INST), DGIST, Daegu 42988, South Korea}
\affiliation{Department of Interdisciplinary Engineering, DGIST, Daegu 42988, South Korea}

\date{5th Nov 2026}

\maketitle

\begingroup
\renewcommand{\thefootnote}{*}
\footnotetext{These authors contributed equally to this work.}
\renewcommand{\thefootnote}{\dag}
\footnotetext{byoungmoo.ann@kriss.re.kr}
\renewcommand{\thefootnote}{\P}
\footnotetext{Responsibility for the preparation of the supplementary information; the co-authors contributed through review.}
\renewcommand{\thefootnote}{\ddag}
\footnotetext{hc2725@gmail.com}
\renewcommand{\thefootnote}{\S}
\footnotetext{minkyung.jung@dgist.ac.kr}
\endgroup

\section*{Supplementary Note 1: Theoretical description}

\subsection*{Classical equation of motion for transmission}

\subsubsection*{Input-output relation}

Using input-output theory, we calculate the transmission spectrum $|S_{21}|^2$ of the circuit shown in Fig. 3(d) illustrating the coupling between the magnon and the resonators. At the port $j$ at which the transmission line couples to the $j$-th resonator, the relation between the input and output fields $a_{j,\text{in/out}}$ to the $j$-th resonator mode

\begin{align}
    a_{1,\text{out}} &= a_{1,\text{in}} - \sqrt{\frac{\kappa_{\text{ex}}}{2}} c_1, \tag{1} \\
    a_{2,\text{out}} &= a_{2,\text{in}} - \sqrt{\frac{\kappa_{\text{ex}}}{2}} c_2, \tag{2} \\
    a_{3,\text{out}} &= a_{3,\text{in}} - \sqrt{\frac{\kappa_{\text{ex}}}{2}} c_3, \tag{3}
\end{align}
$c_j$ is the annihilation operator for the $j$-th resonator mode, 
where $\kappa_{\text{ext}}$ is the external coupling rate to the transmission line. In our case, $c_1 = c_u$,  $c_2 = c_a$, and  $c_3 = c$ are satisfied.
We also have relations among input/output fields like $a_{2,\text{in}} = e^{i\Phi_{21}} a_{1,\text{out}}$ and $a_{3,\text{in}} = e^{i\Phi_{32}} a_{2,\text{out}}$, where $\Phi$ are the phase delays in probe fields due to the spatial separations of each resonator. In our case, $e^{i\Phi_{21}} = e^{i\Phi_{32}} \approx 1$ since the separations among each LER (1--2 mm) are much shorter than the wavelength of the input and output fields (28 mm for frequency of 4.5 GHz). Note that the 1st resonator is coupled to the acoustic ($b$) and optical ($b_o$) magnon modes. $\gamma_{a,o}$ refer to the damping rates of each magnon modes. 

\subsubsection*{Steady state solution}

The steady state equations of motions for the resonators and magnon modes in the frequency domain are given by
\begin{align}
    -i\omega c_u & = -i\left[\omega_{cu} c_u + g_{c} c_a \right] - \frac{\kappa_u}{2} c_u + \sqrt{\frac{\kappa_{\text{ex}}}{2}} a_{1,\text{in}} \tag{4} \\
    -i\omega c_a & = -i\left[\omega_{ca} c_a + g_{c} c_u + g_{c} c \right] - \frac{\kappa_{\text{ex}}}{2} c_u - \frac{\kappa_a}{2} c_a + \sqrt{\frac{\kappa_{\text{ex}}}{2}} a_{1,\text{in}} \tag{5} \\
    -i\omega c & = -i\left[\omega_{c} c + g_{c} c_a + g_{b} b + g_{bo} b_o \right] - \frac{\kappa_{\text{ex}}}{2} c_u - \frac{\kappa_{\text{ex}}}{2} c_a - \frac{\kappa}{2} c + \sqrt{\frac{\kappa_{\text{ex}}}{2}} a_{1,\text{in}} \tag{6} \\
    -i\omega b & = -i\left(\omega_b - i\frac{\gamma}{2}\right) b - i g_{b} c \tag{7} \\
    -i\omega b_o & = -i\left(\omega_{bo} - i\frac{\gamma_o}{2}\right) b_o - i g_{bo} c \tag{8}
\end{align}
$g_{c}$ denotes the coupling among the resonators. We define $g_{b}$ and $g_{bo}$ as the coupling strengths between each magnon mode and auxiliary resonator. By plugging Eqs. (7-8) into Eq. (9), we obtain three coupled equations for three variables $c_u, c_a$, and $c$. Meanwhile, the experimentally observed transmission $S_{21}$ is related with the resonator in-fields by the following equation.

\begin{equation}
    S_{21} = \frac{a_{3,\text{out}}}{a_{1,\text{in}}} \approx 1 - \frac{1}{a_{1,\text{in}}} \sqrt{\frac{\kappa_{\text{ex}}}{2}} (c + c_a + c_u). \tag{9}
\end{equation}

\subsection*{Derivation of Hybridized Decay Rates in the Strong Coupling Regime}

To derive the explicit forms of the decay rates $\gamma^{\pm}$ for the coupled system comprised on the auxiliary cavity and magnon mode, we start with the effective non-Hermitian Hamiltonian
\begin{equation}
H = \begin{pmatrix} \omega_{\text{c}} - i\kappa & g_{b} \\ g_{b} & {\omega}_{b} - i\gamma \end{pmatrix}
\end{equation}
where $\omega_{c}$ ($\omega_{\text{b}}$) and $\kappa$ ($\gamma$) are the resonance frequency and decay rate of the cavity (magnon) mode, respectively, and $g_{\text{b}}$ is the coupling strength. We decompose the Hamiltonian into its Hermitian part $H_0$ and anti-Hermitian part $H_{\text{loss}}$:
\begin{equation}
H = H_0 - iH_{\text{loss}} = \begin{pmatrix} \omega_{c} & g_{b} \\ g_{b} & {\omega}_{b} \end{pmatrix} - i \begin{pmatrix} \kappa & 0 \\ 0 & \gamma \end{pmatrix}
\end{equation}

In the strong coupling regime ($g_{\text{b}} \gg |\kappa - \gamma|$), the eigenmodes of the system are primarily determined by $H_0$. We diagonalize $H_0$ using a rotation matrix $S$:
\begin{equation}
S = \begin{pmatrix} \cos\theta & \sin\theta \\ -\sin\theta & \cos\theta \end{pmatrix}
\end{equation}
where the mixing angle $\theta$ is defined by the relation:
\begin{equation}
\tan(2\theta) = \frac{2g_{\text{b}}}{\omega_{\text{c}} - \omega_{\text{b}}}
\end{equation}
Applying this transformation to $H_0$ yields the hybridized frequencies $\omega_{d\pm}$ as the diagonal elements:
\begin{equation}
S H_0 S^T = \begin{pmatrix} \omega_{d+} & 0 \\ 0 & \omega_{d-} \end{pmatrix}
\end{equation}

Next, we transform the loss matrix $H_{\text{loss}}$ into the hybridized basis:
\begin{equation}
H_{\text{loss}}' = S H_{\text{loss}} S^T = \begin{pmatrix} \kappa\cos^2\theta + \gamma\sin^2\theta & \frac{1}{2}(\gamma - \kappa)\sin(2\theta) \\ \frac{1}{2}(\gamma - \kappa)\sin(2\theta) & \kappa\sin^2\theta + \gamma\cos^2\theta \end{pmatrix}
\end{equation}

The total Hamiltonian in the hybridized basis becomes:
\begin{equation}
H' = S H S^T = \begin{pmatrix} \omega_{d+} - i\gamma^{+} & i\gamma' \\ -i\gamma' & \omega_{d-} - i\gamma^{-} \end{pmatrix}
\end{equation}
Under the condition of strong coupling, the frequency splitting $|\omega_{d+} - \omega_{d-}|$ significantly exceeds the off-diagonal damping terms $\gamma'$. In this limit, the secular approximation allows us to neglect the off-diagonal terms. Consequently, the effective decay rates $\kappa_{\pm}$ of the hybridized modes (CMPs) are given by the diagonal elements of $H_{\text{loss}}'$:
\begin{equation}
\begin{aligned}
\gamma^+ &\approx \kappa\cos^2\theta + \gamma\sin^2\theta \\
\gamma^- &\approx \kappa\sin^2\theta + \gamma\cos^2\theta
\end{aligned}
\end{equation}
This representation provides a physical intuition where the hybridized decay rates are weighted averages of the original constituent decay rates, governed by the mixing angle $\theta$.

\subsection*{Quantum equation of motion}

The total Hamiltonian of the system presented in Fig. 5(a) of the main text, consisting of a superconducting transmon qubit, an auxiliary resonator, and a magnon mode, is given by
\begin{equation}
    H = H_0 + H_{\text{int}} + H_{\text{probe}},
\end{equation}
where $H_0$ represents the free Hamiltonian of each component and $H_{\text{int}}$ describes the coherent energy exchange between them. $H_{\text{probe}}$ describes the probe tone. The dissipation terms will be added in the other form later.
The free Hamiltonian $H_0$, including the transmon's anharmonicity $A_q$, is
\begin{equation}
    H_0 = \omega_q q^\dagger q - \frac{A_q}{2} q^\dagger q^\dagger q q + \omega_{c} c^\dagger c + \omega_b b^\dagger b,
\end{equation}
where $q$, $c$, $b$ ($q^\dagger$, $c^\dagger$, $b^\dagger$) are the annihilation (creation) operators for the transmon, resonator, and magnon modes.
The interaction Hamiltonian is
\begin{equation}
    H_{\text{int}} = g_{b} (b^\dagger c + b c^\dagger) + g_{c} (c^\dagger q + c q^\dagger).
\end{equation}
Here, $g$ is the coupling strength between the magnon and the auxiliary resonator, and $g_c$ is the one between the the auxiliary resonator and the transmon.
Meanwhile, the probe Hamiltonian with strength $\epsilon$ reads
\begin{equation}
    H_{\text{probe}} = i\epsilon(c^\dagger e^{-i\omega t} - c e^{i\omega t} ).
\end{equation}
We then perform a unitary operation $\hat{U}_{\text{rot}}(t)= e^{-i\omega t( \hat{q}^{\dagger}\hat{q}+\hat{c}^{\dagger}\hat{c}+\hat{b}^{\dagger}\hat{b})}$ on $\hat{H}$ to move on to the frame where the Hamiltonian becomes time-independent. In this frame, $H$ is transformed to $H'$ that reads
\begin{equation}
    H' = (\omega_q-\omega) q^\dagger q - \frac{A_q}{2} q^\dagger q^\dagger q q + (\omega_{c}-\omega) c^\dagger c + (\omega_b-\omega) b^\dagger b +  g_{b} (b^\dagger c + b c^\dagger) + g_{c} (c^\dagger q + c q^\dagger) + i\epsilon(c^\dagger- c ).
\end{equation}
To account for the interaction with the environment, we use the Lindblad master equation for the density matrix of the tripartite system $\rho$
\begin{equation}
    \frac{d\rho}{dt} = -i[H', \rho] + \sum_{L} \mathcal{D}[L]\rho
\end{equation}
where the dissipator is defined as $\mathcal{D}[L]\rho = L\rho L^\dagger - \frac{1}{2}\{L^\dagger L, \rho\}$.
In our case, the master equation becomes
\begin{equation}
    \begin{aligned}
        \frac{d\rho}{dt} = -i[H', \rho] & + \kappa_q \mathcal{D}[q]\rho + \kappa_a \mathcal{D}[c]\rho + \gamma \mathcal{D}[b]\rho
    \end{aligned}
\end{equation}
where $\kappa_q$, $\kappa_a$, and $\gamma$ are the dissipation rates of the transmon, auxiliary cavity and magnon modes, respectively.

\section*{Supplementary Note 2: Numerical simulation}

\subsection*{Classical dynamics}

To calculate $S_{21}$, we use \textsf{np.linalg.solve} function of NumPy library. Below, we present the Python script used for the calculation for the tripartite configuration. The parameters are adjusted to emulate the real experiment presented in the main text.

\begin{lstlisting}[language=Python, caption=Python code for calculating the resonator mode amplitudes by solving the coupled linear system.]
import numpy as np

# Unit : 2pi * GHz, Tesla.

g_c      = 0.0055
g_ma     = 0.325
g_mo     = 0.025
omega0_cu = 4.194    # Bare frequency of the unused cavity.
omega0_ca = 4.109    # Bare frequency of the auxiliary cavity.
omega0_c  = 4.015    # Bare frequency of the cavity.
k_ext    = 0.00025 
k_u      = 0.00080   # Damping rate of the unused cavity.
k_a      = 0.00080   # Damping rate of the auxiliary cavity.
k        = 0.00300   # Damping rate of the cavity.
gamma_a  = 0.090
gamma_o  = 0.010

def calculate(omega, B, a_in):
    omega_b = 42 * B
    if 4.5**2 - 150 * B**2 > 0:
        omega_bo = np.real(np.sqrt(4.5**2 - 150 * B**2))
    else:
        omega_bo = 0
        
    omega_1 = omega0_cu - 0.05*B**2 - 0.25*B**4
    omega_2 = omega0_ca - 0.06*B**2 - 0.25*B**4
    omega_3 = omega0_c  - 0.17*B**2 - 0.75*B**4

    # Constructing the 3 x 3 linear system A * c = B
    A = [ 
        [-1j*omega + 1j*omega_1 + k_u/2, -1j*g_c, -1j*g_c],  
        [-1j*g_c + k_ext/2, -1j*omega + 1j*omega_2 + k_/2, -1j*g_c],  
        [k_ext/2 - 1j*g_c, -1j*g_c + k_ext/2, 
         -1j*omega + 1j*omega_3 + k_3/2 
         + 1j*g_bo*(-1j*g_bo/(1j*(omega_bo-omega) + gamma_o/2)) 
         + 1j*g_b *(-1j*g_b/(1j*(omega_b-omega) + gamma_a/2))] 
    ]
    
    B = [np.sqrt(k_ext/2)*a_in, np.sqrt(k_ext/2)*a_in, np.sqrt(k_ext/2)*a_in]

    A = np.asarray(A)
    B = np.asarray(B)

    C = np.linalg.solve(A, B)
    
    return C[0], C[1], C[2]    # c1, c2, c3, respectively
\end{lstlisting}

\subsection*{Quantum dynamics}

To numerically calculate the transmission spectrum and the steady-state populations of the cavity-magnon-transmon system, we utilized the Quantum Toolbox in Python (QuTiP) version 5.1.1, a Python library designed for numerical quantum simulation. The steady-state density matrix $\rho_{\text{ss}}$ is obtained by solving the Lindblad master equation. The following Python script details the construction of the Hamiltonian, the corresponding collapse operators, and how to solve the $\rho_{\text{ss}}$. 

\begin{lstlisting}[language=Python, caption=Python code for alculating the steady-state spectroscopy of the  system using QuTiP.]
import numpy as np
from qutip import *

# Unit : 2pi * GHz, Tesla.

def spectroscopy(omega_p_list, eps, B):
    
    Nq = 3
    Nc = 5
    Nm = 5
    omega_q      = 4.15
    omega_c      = 4.0 - 0.05*B**2 - 0.25*B**4
    omega_b      = 42 * B
    Anh     = 0.2
    
    kappa_q   =  0.0010
    kappa_c   =  0.0025
    gamma     =  0.0900

    g_b       =  0.40
    g_c       =  0.02
    
    q = tensor(destroy(Nq), qeye(Nc)    , qeye(Nm)   )
    c = tensor(qeye(Nq)   , destroy(Nc) , qeye(Nm)   )
    b = tensor(qeye(Nq)   , qeye(Nc)    , destroy(Nm))

    num_q = q.dag()*q
    num_c = c.dag()*c
    num_b = b.dag()*b

    transmission=[]
    
    nth_q = 0.0
    nth_c = 0.0
    nth_b = 0.0
    
    H0 = (omega_q)*num_q + (omega_c)*num_c + (omega_b)*num_b \
         - 0.5*Anh*q.dag()*q.dag()*q*q \
         + g_b*(c.dag()*b + c*b.dag()) \
         + g_c*(c.dag()*q + c*q.dag()) \
         - 1j*eps*(c.dag() - c) 
    
    c_ops = []
    c_ops.append(np.sqrt(kappa_q*(nth_q+1))*q      )
    c_ops.append(np.sqrt(kappa_q*(nth_q  ))*q.dag())
    c_ops.append(np.sqrt(kappa_c*(nth_c+1))*c      )
    c_ops.append(np.sqrt(kappa_c*(nth_b)  )*c.dag())
    c_ops.append(np.sqrt(gamma*(nth_b+1))*b      )
    c_ops.append(np.sqrt(gamma*(nth_b)  )*b.dag())
    
    transmission = []
    magnon_num   = []
    photon_num   = []
    
    for omega_p in omega_p_list:                               
        
        H  = H0
        H -= omega_p *num_q
        H -= omega_p *num_c
        H -= omega_p *num_b
        
        rho_ss = steadystate(H, c_ops)
        transmission.append((expect(c, rho_ss)))
        magnon_num.append((expect(num_b, rho_ss)))
        photon_num.append((expect(num_c, rho_ss)))

    return np.array(transmission), np.array(magnon_num), np.array(photon_num)

\end{lstlisting}

\section*{Supplementary Note 3: Additional experimental data}

Fig. S1(a–b) shows the experimentally observed (a) and numerically calculated (b) transmission of the additional device, which has the same configuration as in Fig. 3(a) in the main text. The observed and calculated spectra agree excellently with each other. In the numerical calculation, we use $g_b/2\pi = 270\text{ MHz}$, $g_{bo}/2\pi = 25\text{ MHz}$, $\kappa/2\pi = 3.5\text{ MHz}$, $\kappa_{a,u}/2\pi = 1.5\text{ MHz}$, $\kappa_{ex}/2\pi=0.35\text{ MHz}$, $g_c/2\pi = 10\text{ MHz}$ and $\gamma/2\pi = 150\text{ MHz}$ by comparing the observation and calculation. We use $\omega_b/2\pi = 42 \text{ GHz/T} \times |\vec{B}| $ in the calculation.
Fig. S1(c--d) provide magnified views near the secondary anti-crossings. The clear avoid-crossing represents the strong and resonant coupling between the upper CMP and target cavity modes. The upper branch of the avoid-crossing is invisible as they form a dark state to the feedline. The extracted coupling strength between them is $g_c^+/2\pi \approx 9.7\text{ MHz}$.

The estimated $\gamma^+$ with respect to $|\vec{B}|$ is presented in Figs.~S1(e) and S1(f) as solid lines. Around the resonance ($|\vec{B}| \approx 0.09\,\mathrm{T}$), $\gamma^+$ is an order of magnitude smaller than $\gamma$. 
Assuming an ideal scenario where the intrinsic losses of the  cavity are negligible (dashed line), a lower $\gamma^+$, and consequently, a larger cooperativity can be achieved for the given $|\vec{B}|$. The circles in Fig.~S1(f) points when the secondary anti-crossing becomes on-resonant.

\begin{figure*}[htbp]
    \centering
    \includegraphics[width=0.7\linewidth]{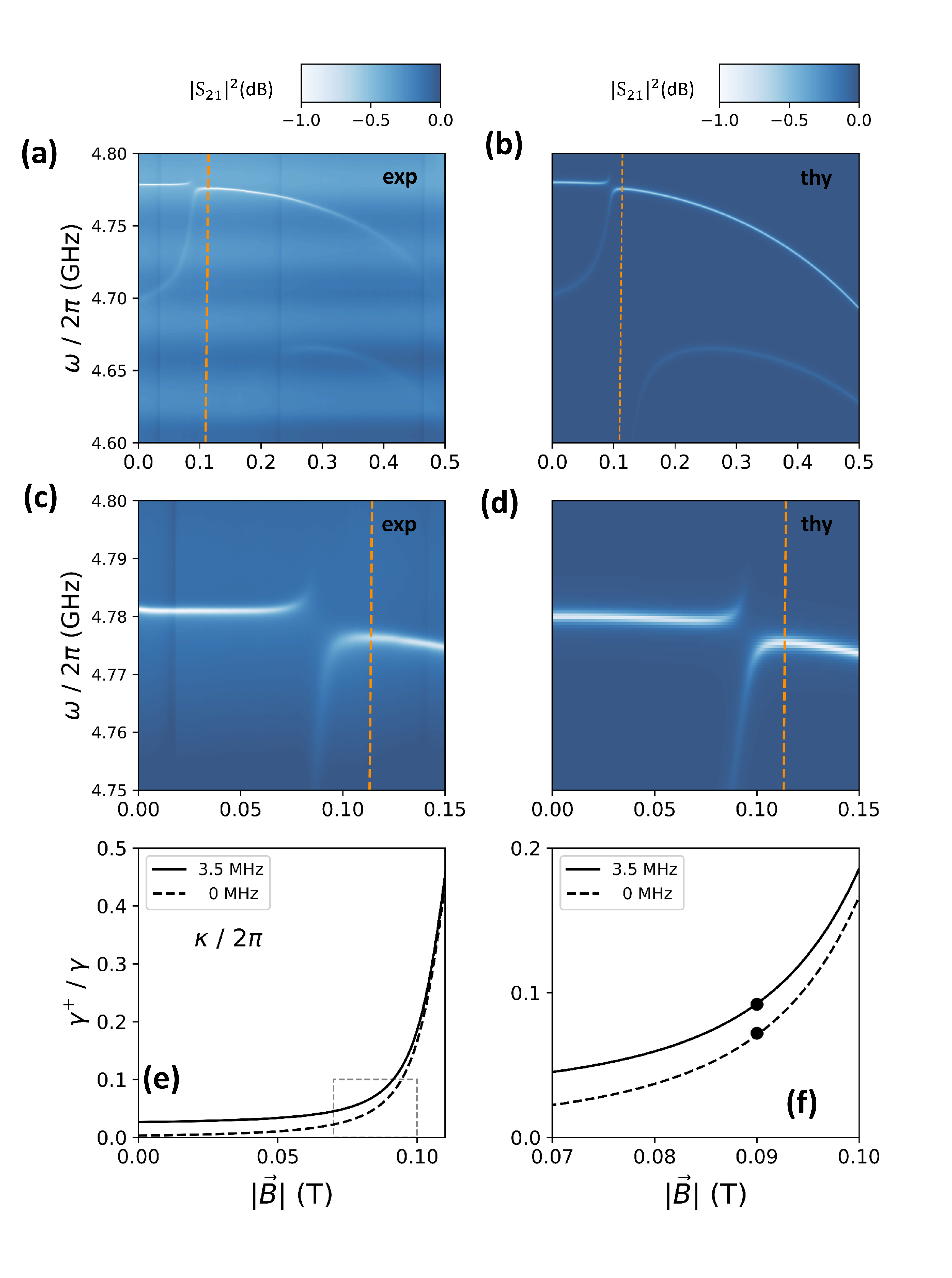}
    \caption{\textbf{Additional data.} 
    \textbf{(a)} The observed transmission spectrum $|S_{21}|^2$ of the additional device with the same configuration in Fig. 3(a) of the main text. $|S_{21}|^2$ is plotted with respect to probe frequency ($\omega/2\pi$) and the magnitude of the in-plane magnetic field ($|\vec{B}|$) at the temperature $T = 1.7$~K.
    \textbf{(b)} Calculated transmission spectrum $|S_{21}|^2$, which exhibits excellent agreements with the observation.
    \textbf{(c--d)} Magnification near the secondary anti-crossing. The orange lines in (a--d) is the magnon resonance frequency $\omega_b$.
    \textbf{(e--f)} Estimated $\gamma^+/\gamma$ with respect to $|\vec{B}|$ from the values used in the numerical simulation in \textbf{(b)} and \textbf{(d)}. The circles indicate the points when the secondary aviod crossing is on-resonant.}
    \label{fig:resonant_interaction}
\end{figure*}